\documentclass[conference,10pt]{IEEEtran}
\let\ifpaper\iftrue
\ifCLASSOPTIONcompsoc
  \usepackage[nocompress]{cite}
\else
  \usepackage{cite}
\fi
\usepackage{amsmath, amssymb}
\usepackage[english]{babel}
\usepackage{comment}
\usepackage[mathscr]{euscript}
\usepackage{color}
\usepackage{url}
\usepackage{textcomp}
\usepackage{clrscode}
\usepackage{times}
\usepackage{xspace}

\usepackage[utf8]{inputenc}

\usepackage[caption=false]{subfig}
\usepackage{commath}
\usepackage{graphicx,bm}
\usepackage{verbatim}
\providecommand{\ie}{{i.e.,} }
\providecommand{\eg}{{e.g.,} }

\providecommand{\mypara}[1]{\smallskip\noindent\emph{#1} }
\providecommand{\myparab}[1]{\smallskip\noindent\textbf{#1} }

\newcommand{\sysname}{Rosella\xspace}

\newcommand{\tv}{\mathrm{TV}}

\newcommand{\bD}{\mathbf{D}}

\newcommand{\bvv}{\mathbf{v}}
\newcommand{\buv}{\mathbf{u}}

\newcommand{\sqq}{\mathrm{SQ}}
\newcommand{\llp}{\mathrm{LL}}

\usepackage{enumitem,kantlipsum}
\newcommand{\E}{\mathrm{E}}
\usepackage{fancyhdr}
\usepackage{amsthm}
\usepackage{etoolbox}

\newtheorem{definition}{Definition}
\AfterEndEnvironment{definition}{\noindent\ignorespaces}

\newtheorem{proposition}{Proposition}

\newtheorem{lemma}{Lemma}

\newtheorem{example}{Example}

\AfterEndEnvironment{proposition}{\noindent\ignorespaces}
\AfterEndEnvironment{example}{\noindent\ignorespaces}
\AfterEndEnvironment{problem}{\noindent\ignorespaces}
\AfterEndEnvironment{theorem}{\noindent\ignorespaces}

\begin{document}
\title{\sysname: A Self-Driving Distributed Scheduler for Heterogeneous Clusters}

\author{
Qiong Wu and Zhenming Liu\\
\quad  William \& Mary \;\;\; 
}

\maketitle
\vspace{-3mm}

\begin{abstract}

Large-scale interactive web services and advanced AI applications make sophisticated decisions in real-time, based on executing a massive amount of computation tasks on thousands of servers. Task schedulers, which often operate in heterogeneous and volatile environments, require high throughput, i.e., scheduling millions of tasks per second, and low latency, i.e., incurring minimal scheduling delays for millisecond-level tasks. Scheduling is further complicated by other users' workloads in a shared system, other background activities, and the diverse hardware configurations inside datacenters. 

We present \sysname, a new self-driving, distributed approach for task scheduling in heterogeneous clusters. \sysname automatically \emph{learns} the compute environment and \emph{adjusts its scheduling policy} in real-time. The solution provides high throughput and low latency simultaneously because it runs in parallel on multiple machines with minimum coordination and only performs simple operations for each scheduling decision. Our learning module monitors total system load and uses the information to dynamically determine optimal estimation strategy for the backends' compute-power. \sysname generalizes power-of-two-choice algorithms to handle heterogeneous workers, reducing the max queue length of $O(\log n)$ obtained by prior algorithms to $O(\log \log n)$. We evaluate \sysname with a variety of workloads on a 32-node AWS cluster. 
Experimental results show that Rosella significantly reduces task response time, and adapts to environment changes quickly.

\end{abstract}

\section{Introduction}
The recent explosion of artificial intelligence (AI) and machine learning significantly altered the compute workloads in backend data centers~\cite{liu2019near}. Users today expect real-time delivery of highly intelligent services to their devices~\cite{cluster21-weighttransfer,8665577,wu2020bats}. Search engines continuously predict search queries and refresh search results on browsers within a few tens of milliseconds, requiring the processing of enormous tiny tasks on thousands of machines~\cite{Yu2011,Dean2013}. Virtual and augmented reality devices continuously analyze video and render graphics based on the analysis results. The emerging class of advanced AI applications (e.g., autonomous vehicles~\cite{wu2015early}, assets pricing~\cite{wu2019adaptive,wu2019deep} and robotics) need to perform many simulations in strict timing requirements to determine the next action when interacting with physical or virtual environments~\cite{Silver2016}.

It becomes prohibitively expensive to process such workloads in dedicated compute clusters. A costly single dedicated GPU-server can process only one or two video streams~\cite{aws-dl}, yet consumer products or surveillance solutions require analyzing thousands of data streams simultaneously. Therefore, recent data-intense systems often attempt to tape in highly volatile, lower-cost computing sources.  For example, AWS scales to the computation demands while also provides a reduced-cost option by leasing its under-utilized boxes (e.g.~T instances, or spot instance bidding~\cite{aws-spot,aws-burst}). However, three challenges must be solved to schedule applications on such systems efficiently at scale: 


\myparab{High-throughput and low latency requirement.} Task schedulers are now required to provide high throughput as they need to schedule millions of tasks per second for these applications~\cite{ousterhout2013sparrow,ray-osdi}. At the same time, 
scheduling needs to be low latency because the tasks require responses at a millisecond level~\cite{reddi2020mlperf}.

\myparab{Heterogeneous environments.} 
Task schedulers operate in environments composed of  CPUs, GPUs, FPGAs, and specialized ASICs~\cite{liu2020gpus,ibrahim2019analyzing,lin2018gpu,liu2016lightweight,liu2018architectural}. Administrators may rent servers from public clouds (AWS, Azure, etc.) and markets (AWS marketplace).  Different types of servers may be rented to minimize their changing prices and cost efficacy. Organizations using private clouds may host servers of different generations to gradually upgrade servers. 
Advanced cross-platform machine learning frameworks (e.g., TensorFlow~\cite{abadi2016tensorflow}) can also be executed on heterogeneous boxes such as smartphones, consumer-grade PCs, high-end GPUs, and other devices.

\myparab{Unknown and evolving compute-power.} 
Workers' performances are often time-varying in practice. For example, multiple groups in a large organization share the same clusters.  The computing power of servers controlled by a different group's scheduler may drop when an adjacent group launches a large batch of jobs~\cite{greenberg2015building}. 
For example, instances offered by AWS come from residual/under-utilized resources, and the compute-throughput fluctuate~\cite{aws-burst}.

\begin{figure}[t!]
\centering

\includegraphics[scale=0.5]{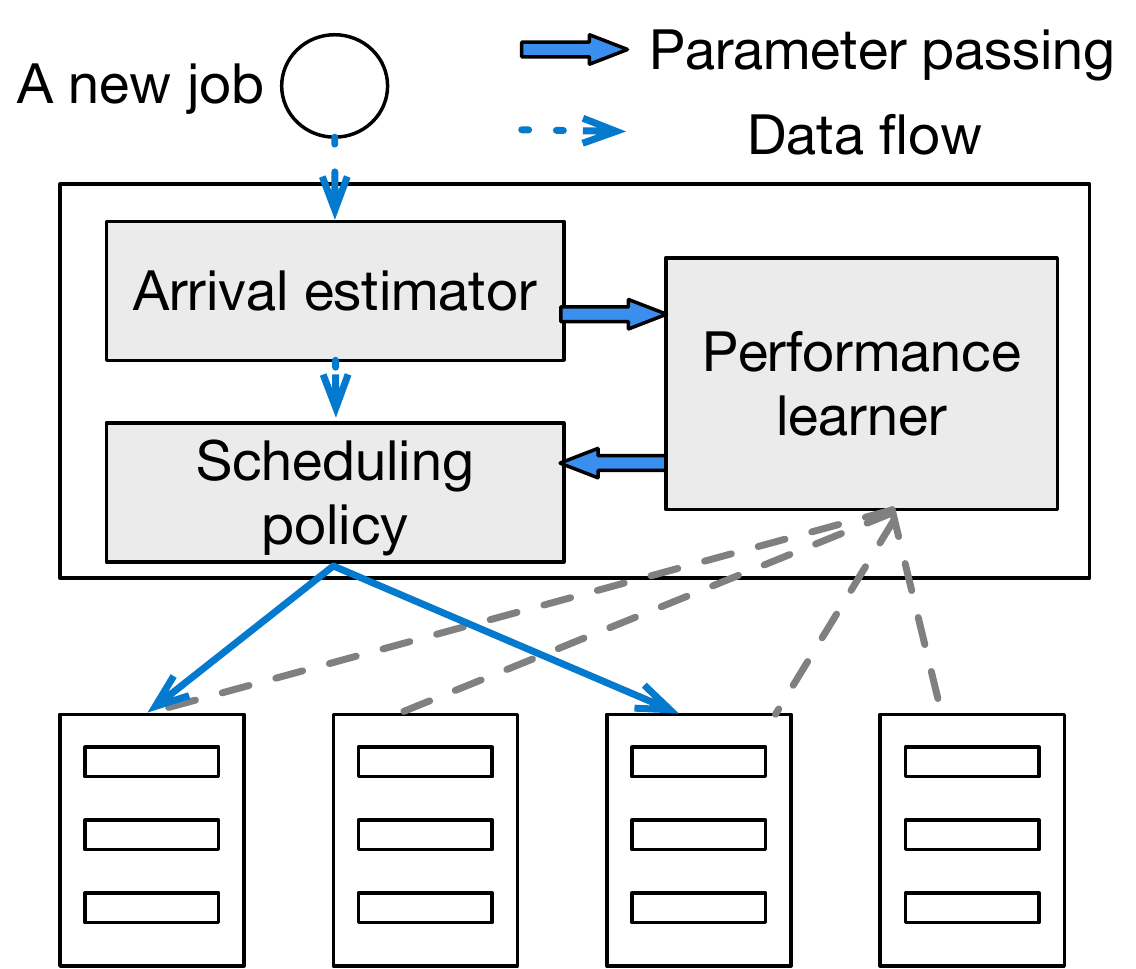}

  \caption{
  {The architecture of \sysname: it continuously learns backend workers' performance and adjusts its scheduling policy.}}
    \label{fig:architecture}
    \vspace{-2mm}
\end{figure}

We present \sysname, a high-throughput and low-latency self-driving scheduler for heterogeneous systems.
\sysname continuously adjusts its policy as the workers' compute power fluctuates in a \emph{self-driving} way. 
\sysname learns the workers' processing power and acts on the learned parameters  simultaneously. Specifically, \sysname possesses the following salient features (see also Fig.~\ref{fig:architecture}):


\vspace{-1mm}
\myparab{1. Efficiently learning the parameters.}
\sysname efficiently estimates the processing power of each worker. \sysname's learning-time scales inverse-proportional to the load ratio, and logarithmically to the number of servers. Both dependencies are essentially optimal, making the scheduler highly scalable. For example, when the number of servers doubles, learning time only increases by a constant unit of times.

\vspace{-1mm}
\myparab{2. Heterogeneity-aware schedulers.}
We unify two major scheduling techniques in our job-allocation algorithm. The first is the so-called \emph{proportional sampling strategy}~\cite{gandhi2015halo}. When a new task arrives, the scheduler chooses a worker according to a multinomial distribution so that the probability that the $i$-th worker is chosen is proportional to its compute-power, e.g., if the $i$-th worker is five times faster than the $j$-th worker, the $i$-th worker is five times more likely to be chosen. The second is the 
\emph{power-of-two-choices~\cite{Mitzenmacher00thepower}} (PoT) strategy. When a new task arrives, we execute the proportional sampling algorithm twice to obtain two candidate workers and assign the task to the worker with the shorter queue.

In summary, we make the following contributions:

    $\bullet$  We analyze the root causes why prior scheduling algorithms fail to provide high-throughput and low latency in a heterogeneous cloud system. 
    
    $\bullet$ We propose an algorithm to schedule jobs to a cluster with different compute-power workers. Compared to prior work, our scheduling algorithm reduces the worst-case queue length from $O(\log n)$ to $O(\log \log n)$.  
    
    $\bullet$ We implement our algorithm on top of Spark~\cite{spark} scheduler. Extensive experiments of real-life workloads on a 32-node AWS cluster demonstrate that \sysname~significantly outperforms a state-of-the-art scheduler~\cite{ousterhout2013sparrow} by $65\%$ in response time, and also is robust against various workloads in the dynamic and heterogeneous cluster.

\section{Problem Setting}

We consider a distributed system that consists of $n$ workers. 
Jobs may contain one or more tasks. 
When a new job arrives, the scheduler can probe a certain number of workers and decide how the tasks should be assigned to the workers based on their queue length. 
$\lambda$ is total arrival rate. The compute power of the workers is heterogeneous. The processing/service rate for worker $i$ is $\mu_i$.
Let $\mu = \sum_{i \leq n} \mu_i$ be \emph{the system's the total processing power}, and $\alpha = \lambda / \mu$ be the system's \emph{load ratio}. Both $\mu_i$ and $\lambda$ can change over time to reflect volatility in the system. 
We aim to design a simple scheduling algorithm that simultaneously optimizes the following metrics: 

\mypara{1. Response time.} When a job arrives, how much time does the system need to process the job? 

\mypara{2. Learning and recovery time.} When a system  experiences a shock (many $\mu_i$s changed), the scheduler does not have accurate estimates of the service rates of the workers ($\mu_i$s). How much time does the system need to re-learn the $\mu_i$s, and once learned, how much time does the system need to recover (e.g., handling the backlogs produced by the scheduler using inaccurate estimates of $\mu_i$s)?

\section{Motivation} 

Learning and scheduling algorithms have been widely studied~\cite{smith1978new,Mitzenmacher00thepower,McDiarmid05,moseley2011scheduling,lin2013joint,xie2015power,ying2017power,258866,258870,humphries2021case,kaffes2019centralized}, but none are directly applicable in our setting. Below, we review the commonly used scheduling algorithms and learning algorithms and explain why they are not applicable.

\begin{figure}
     \centering
     \subfloat[Uniform algorithm\label{subfig-1:dummy}]{%
       \includegraphics[width=0.3\textwidth]{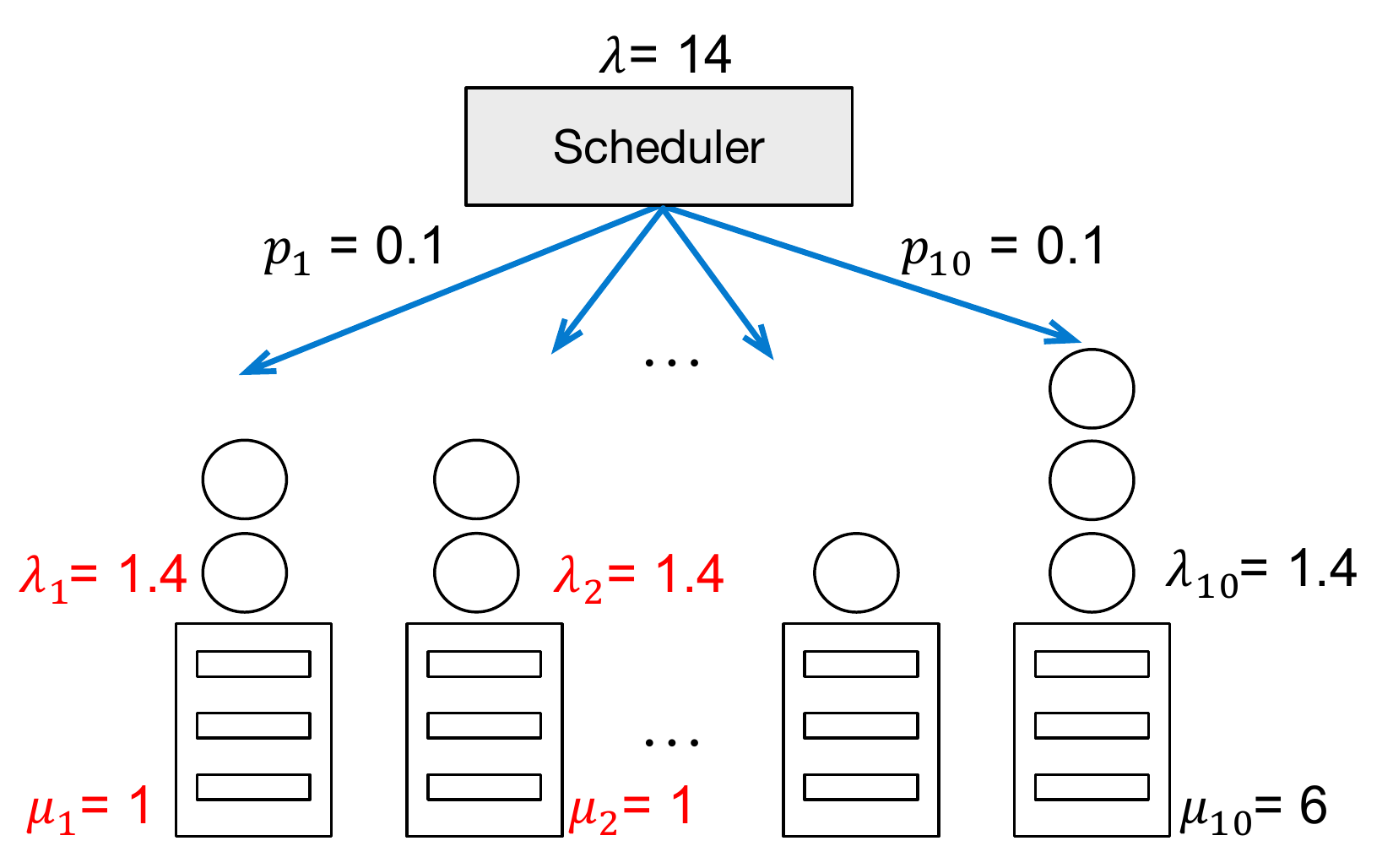}
     }
     \hfill
     \centering
     \subfloat[PoT algorithm\label{subfig-2:dummy}]{%
       \includegraphics[width=0.3\textwidth]{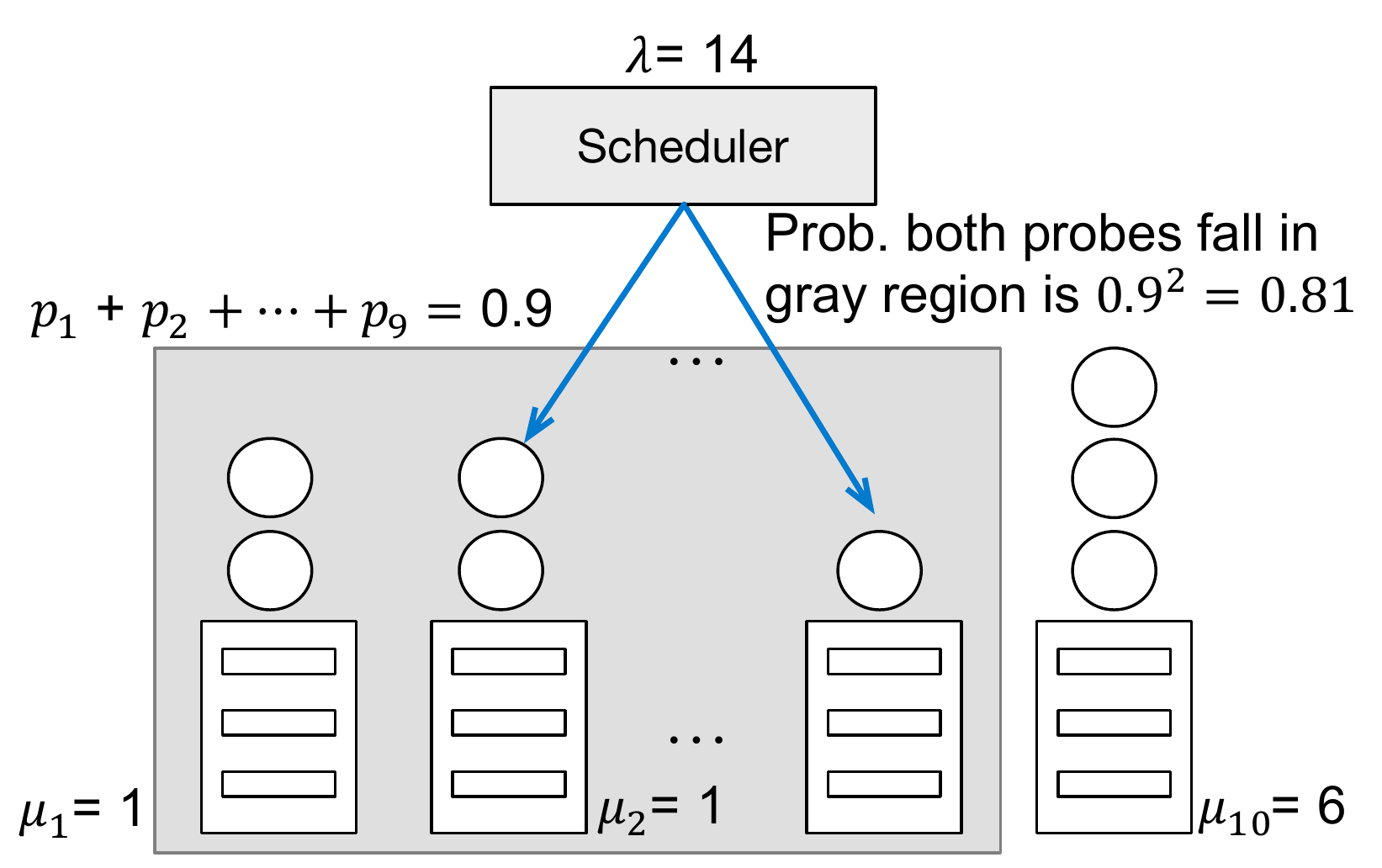}
     }
     \caption{Prior scheduling policies work poorly in a heterogeneous cluster. Uniform algorithm: $\mu_1 = 1$ while $\lambda_1 = 1.4$. implying that in the long run, worker 1 is non-stationary. PoT algorithm: with probability 0.81, workers 1 to 9 receive a job. The aggregate arrival rate to workers 1 to 9 is $14 \times 0.81 =  11.34$, but the total processing rate of workers 1 to 9 is only 9, implying that in the long run workers 1 to 9 are non-stationary.There are 10 servers. $\mu_1 = \mu_2 = \dots = \mu_9 = 1$. $\mu_{10} = 6$. $\lambda = 14$.
     }
     \label{fig:blowup}
    \vspace{-6mm}
   \end{figure}

\subsection{Scheduling Algorithms}

\textbf{Uniform algorithm.} When a new job arrives, the scheduler uniformly chooses a worker to serve it~\cite{Mitzenmacher2005}. When all $\mu_i$s are uniform, the system consists of $n$ independent standard queues. At each time unit, each queue receives $\lambda / n$ tasks and can process $\mu / n$ tasks. All queues can process more tasks than they receive and thus all of them are stationary.  The expected length of the largest queue is $O(\log n)$. 
The uniform algorithm fails work when the service rates are different for different servers, i.e., by assigning the same amount of jobs to each server, the faster servers are underloaded and slower servers are overloaded. 

\begin{example}\label{example:uniform}\noindent\emph{
Fig.~\ref{fig:blowup}a shows the example of uniform algorithm does not work. Assume there are 10 workers. The service rates for workers 1 to 9 are 1. The service rate for worker 10 is 6. The arrival rate is $14$. The uniform algorithm assigns 10\% of jobs to worker 1 (i.e., $\lambda_1 = 1.4$) but $\mu_1 = 1$, implying that in the long run, worker 1 needs to process more tasks than its capacity.}

\end{example}

\myparab{Power of two choices (PoT).} When a new task arrives, the scheduler probes two random workers and assigns the new job to the worker with shorter queue. The PoT algorithm has improved worst-case queue length when the workers are homogeneous  \ie with high probability the largest queue length is $O(\log \log n)$~\cite{Mitzenmacher00thepower}. 
However, the PoT algorithm suffers the same problem where slower workers are overloaded in the heterogeneous environment.


\begin{example}\emph{
Fig.~\ref{fig:blowup}b shows the example of PoT does not work.
Using the same configuration in Example~\ref{example:uniform}, With probability $0.9 \times 0.9= 0.81$, the scheduler selects 2 slow workers so one of them will process the job. On average there are $14 \times 0.81 = 11.34$ jobs arriving at the slow workers, but the slow workers' total processing power is only $9$, implying that in the long run the slow workers need to process more jobs than their capacity.  
}

\end{example}

\myparab{Heterogeneity-aware load balancing.} 
Recent attempts address the heterogeneity issue via assigning more jobs to more powerful workers~\cite{gandhi2015halo}. However, they impractically assume accurate knowledge of the servers' processing powers and  the processing powers do not change over time.

\subsection{Learning Algorithms}\label{sec:priorlearning}


\noindent \textbf{Explore-exploit paradigm.}
One reliable way to estimate $\mu_i$ is to compute the average processing time of the sufficient recent tasks. The explore-exploit paradigm arises because we do not want to assign many jobs to slow workers (i.e., exploit more powerful ones) and we also want to closely track the slower workers' processing power so that we can use it when it becomes faster (explore the weaker ones). 

The explore-exploit paradigm has been widely studied (e.g., multi-arm bandit problems~\cite{bubeck2012regret}), but the solutions mostly focus on the so-called regret bound. Regret analysis assumes that the regrets are memoryless, (i.e., a wrong decision is only penalized once). However, in our setting, a wrong decision may have a long-lasting effect (i.e., an adversarial impact on all subsequent load-balancing decisions). 

\mypara{Our learning objective.}  Instead of minimizing the regret, our learning and schedule algorithm simultaneously must \emph{1. Efficiently learn the processing power.} When the system is cold-started or has recently experienced a shock, the algorithm needs to efficiently (re)-learn the workers' processing power; \emph{2. Rapidly converge to the stationary distribution.} When the algorithm re-learns the workers' processing power, the system rapidly converges to the stationary distribution (i.e., efficiently handles the backlogs from using inaccurate/old processing power estimates); and \emph{3. Be robust against estimation errors.} The scheduler must work well in the presence of small estimation error.

\section{System Design}

This section explains \sysname's architecture, and the operational details. Fig.~\ref{fig:architecture} shows the architecture and the three main components: arrival estimator (Sec.~\ref{arrival_estimator}), scheduling policy (Sec.~\ref{sec:potalgo}) and performance learner (Sec.~\ref{performance_learner}).
When a job arrives, it goes to the arrival estimator, which estimates and updates the arrival rate $\lambda$ of the system. Next, the job goes to the scheduling policy, which uses the estimates of $\mu_i$ provided by the performance learner to choose the appropriate worker. 
The performance learning, which operates in the background, continuously maintains current estimates of each worker's processing power. It takes the estimate $\lambda$ as input and determines 
how to communicate with the workers.

\myparab{Technical challenges.}
\mypara{1. Power-of-two generalization.} Recall that a classical PoT algorithm uniformly samples two workers, and uses the one with lighter loads to process the new job. Generalization of the PoT algorithm requires us to use non-uniform sampling (see Sec.~\ref{sec:potalgo} for further discussions), and redefine the rule of choosing a worker because both of the two policies below are plausible (see \eg~\cite{foss1998stability,foley2001join,bramson2010randomized,bramson2012asymptotic}): 

\begin{enumerate}[leftmargin=0.7cm]
\item The \emph{join the shortest queue} policy ($\sqq$): assigns the incoming job to the queue with the shorter length. 
\item The \emph{join the least loaded queue} policy ($\llp$): assigns the incoming job to the queue with shorter waiting time. 
\end{enumerate}

Sec.~\ref{sec:potalgo} discusses our design choice on these policies.

\mypara{2. The explore-exploit paradigm and more jobs are better.}
Our performance learning component must balance the tradeoffs between estimating slow workers (that could potentially become fast) and assigning more jobs to fast workers. While this is a classical explore-exploit paradigm extensively studied in multi-arm bandit problems, one key difference is that in multi-arm bandit problems, a scheduler needs to passively perform the explore operations (e.g., only when a job arrives can the scheduler use it to explore) the workers' performance; whereas our algorithm can actively explore servers' performance by \emph{creating new jobs.}

Optimizing the learning performance involves carefully controlling the number of jobs that need to be to be created for the purpose of exploration. Creating too few jobs will not accelerate the learning process, while creating too many jobs will slow down the whole system. 

\subsection{The arrival estimator}\label{arrival_estimator}

This component estimates $\lambda$, using the mean interarrival time for the last $S$ jobs as the estimation of $1/\lambda$. Here, $S$ is a hyperparameter. When $S$ is large, the estimate of $\lambda$ is more accurate, but the system reacts more slowly to the change of worker speeds. When $S$ is small, the estimate of  $\lambda$ is less accurate, but the system reacts more rapidly to the environment changes. 

\subsection{The scheduling policy}\label{sec:potalgo}

The scheduling policy component (Fig.~\ref{fig:pss}) has access to estimates $\hat \mu_i$ from the performance learning component and schedule the jobs.
Our policy deviates from the classical PoT algorithms in two major ways.

\myparab{1. Proportional sampling schedule (PSS).}
To circumvent heterogeneity of the workers, our approach probes faster workers with higher probability. 
Let $p_i = \hat \mu_i / (\sum_{i \leq n}\hat \mu_i)$. The proportional sampling procedure samples a worker from a multinomial distribution $(p_1, p_2, ..., p_n)$. When the $\hat \mu_i$'s are accurate, workers behave like independent queues under proportional sampling with high probability that the maximum queue length is $O(\log n)$.

\myparab{2. Power-of-two-choices with $\sqq$.} 
We integrate PoT techniques to further reduce the maximum queue length, \ie we use PSS to choose two workers and place the new job to the better one. We can use $\sqq$ or $\llp$ policy. 
\sysname uses $\sqq$ since it avoids using slower servers until too many jobs are waiting at the faster server:

\begin{example}\label{example:scaled}\emph{A system consists of $n = \mu + 1$ worker, where $\mu \gg 1$ is an integer. Worker 1's processing rate is $\mu$ and the other servers'  processing rate is 1. The total processing rate is $2\mu$, and the first server is substantially faster. The arrival rate of the jobs is $1.5\mu$.
}

\end{example}

The probability that worker 1 is chosen as a candidate is $1 - \left(\frac 1 2\right)^2 = \frac 3 4$. When worker  1 is chosen as a candidate, it will pick up the new job if it has less than $\mu - 1$ jobs because the expected processing time for the other workers is at least $1$ (even assuming the other candidate's queue is empty), whereas the expected processing time of worker 1 is $< 1$ when it has less than $\mu - 1$ jobs. 

Assume that the system starts with empty queues at time $0$. In the beginning, the arrival rate to worker 1 is $1.5\mu \times \frac 3 4 = \frac 9 8$ while its processing rate is $\mu$. Thus, the queue quickly builds up until its length hits $\mu - 1$. The queue will not shrink much afterward because as worker 1's queue decreases, worker 1 will attempt to pick up more jobs. Thus, in the stationary state, the length of queue 1 is around $\mu - 1$, and the expected waiting time for a job at worker 1 is $(\mu - 1 + 1)/\mu = 1$, \emph{which is as slow as the other slow servers.}

In general, more jobs will be congested at the faster workers; all the workers could be as slow as the slowest server. 
In the $\sqq$ policy, however, slower workers will be utilized before faster servers become too full, alleviating the congestion problem. 

\begin{figure}
\begin{codebox}
\Procname{$\proc{PPoT-Scheduling-policy}(s_i)$}
\li \Comment $s_i$ is the $i$-th job.
\li Let $p_i = \frac{\hat \mu_i}{\sum_{i \leq n}\hat \mu_i}$ 
\li Let $\vec p = (p_1, \dots , p_n)$. 
\li $j_1, j_2 \gets \mathrm{multinomial}(\vec p)$. 
\li \Comment let $q_i$ be the length of queue $i$.
\li $j^* = \arg \min_{j \in \{j_1, j_2\}}\{q_i\}$. 
\li \Comment place the job at the $j^*$-th server. 

\end{codebox}
\vspace{-2mm}
\caption{Pseudocode for our proportional-sampling+PoT scheduling.}
\label{fig:pss}
\end{figure}

\subsection{Performance learner}\label{performance_learner}

The performance learner, which operates in the background,  continuously maintains current estimates of each worker's processing power. It takes the estimate of $\lambda$ as input and uses it to determine how often to communicate with the workers.   The performance learner actively generates new jobs and assigns them to the workers (see $\proc{Learner-Dispatcher}$ in Fig.~\ref{fig:learner}). The jobs serve as benchmarks to estimate the workers' processing powers.

\sysname generates the benchmark jobs according to a Poisson process with parameter $c_0(\bar \mu - \hat \lambda)$, where $\bar \mu$ is the minimum guaranteed service throughput, $\hat \lambda$ is the estimate of arrival rates, and $c_0$ is a small constant (say 0.1). Generating jobs at this frequency ensures that we optimally monitor all resources in the cluster while not jamming the system and slowing down the processing of other jobs. The benchmark jobs have low priorities, which will not be executed if other ``real'' jobs are waiting in the worker. 

\mypara{Choosing benchmark jobs.} The benchmark jobs shall resemble recent workloads. For example, they can be replicates of the most recent queries at the frontend. 

\mypara{Learning.} When worker $i$ completes computation of a job, it will communicate with the performance learner to update its estimate $\hat \mu_i$ (see $\proc{Learner-aggregate}$ in Fig.~\ref{fig:learner}). The estimate is based on computing the average processing time of the last $L$ jobs, where $L = \Theta(\frac{\log(1/n)}{(1-\hat \alpha)^2})$, and $\hat \alpha = \hat \lambda / \bar \mu$ is the estimated load ratio. The historical window length depends on the load ratio $\alpha$ and total number of jobs $n$ for the following reasons: 
When  $\alpha$ is small, there are sufficient residual compute resources so we can afford to have sloppy estimates. Therefore, the size of the historical window shrinks when $\alpha$ decreases. There is also a dependency on $\log 1/n$ because we need to use the standard Chernoff/union bound techniques to argue that all estimates are in reasonable quality (see Sec.~\ref{sec:analysis} for the analysis).

When a worker is slow, it may take time to collect the statistics over its most recent $L$ tasks. Therefore, we set a waiting time cut-off, i.e., if we cannot estimate $\mu_i$ in $(1+\epsilon)L/\mu^*$ time (the variables in the Fig.~\ref{fig:learner}), we set the estimate as 0, effectively treating the small worker as dead.

\begin{figure}
\begin{codebox}
\Procname{$\proc{Learner-Dispatcher}$}
\li \Comment Dispatches benchmark jobs to workers
\li Sample $t_i \sim \mathrm{Poisson}\left(0.1(\bar \mu - \hat \lambda)\right)$. 
\li At time $t_i$: $j \gets \mathrm{Uniform}([n])$.
\li Assign a low priority job to worker $j$. 
\end{codebox}

\begin{codebox}
\Procname{$\proc{Learner-aggregate}(\mbox{worker } i)$}
\li \Comment Communicate with
performance learner
\li $\hat \alpha \gets \hat \lambda/\bar \mu$, $\epsilon = \frac{3}{10}(1-\alpha)$, $\mu^* = (1-\hat \alpha)/10$.  
\li $L \gets \frac{c_1}{\epsilon^2}\log(1/n)$ for some constant $c_1$. 
\li Let $\hat q_i$ be the average processing time 
\li \quad for the most recent $L$ jobs
\li \If cannot measure $\hat q_i$ in $(1+\epsilon)L/\mu^*$ time, $\hat \mu_i = 0$.
\li \Comment the workers too slow
\li \Else  $\hat \mu_i = (1-\epsilon)1/\hat q_i$  \End
\li Report $\hat \mu_i$ to the performance learner 
\end{codebox}
\caption{Pseudocode for the performance learner.}
\label{fig:learner}
\vspace{-2mm}
\end{figure}

\section{Analysis of the Algorithms}\label{sec:analysis}

\myparab{Our model.} 
We consider a distributed system that consists of $n$ workers/servers, each connected to a scheduler. Jobs arriving to the scheduler may contain one or more tasks. When a new job arrives, the scheduler probe a certain number of workers and decide how  to assign tasks based on workers' queue length. 
For simplicity, our theoretical model focuses on the case of one job containing one task (but our evaluation will consider the general case). The arrivals of jobs follows a Poisson distribution with parameter $\lambda$. Processing/compute power of the workers is heterogeneous. The service time for a job assigned to worker $i$ follows an exponential distribution with parameter $\mu_i$. 
Let $\mu = \sum_{i \leq n} \mu_i$ be \emph{the system's the total processing power}, and $\alpha = \lambda / \mu$ be the system's \emph{load ratio}. 
Both $\mu_i$ and $\lambda$ can change over time to reflect volatility in the system. 
$\bar \mu$, which is always larger than the job arrival rate. 



\myparab{Discrete-time counterpart.} 
We use a standard way to couple our continuous time model with a discrete time model~\cite{ross2014introduction}. The discrete time model proceeds in rounds. One of the following events occurs in each round: (i) With probability $\lambda / (\lambda + \mu)$, one new job arrives, and (ii)  With probability $\mu_i / (\lambda + \mu)$, one processing event happens at worker $i$ as follows: if the worker's queue contains one or more jobs, the the oldest job is processed first. Otherwise, nothing happens. 
One round in the discrete time model corresponds to a ``jump'' event in the continuous time model. Since we have coupled models (e.g., a convergence result for one often implies a similar result for the other), our analysis can switch between them to deliver an intuitive analysis.

\vspace{-1mm}
\myparab{Main results.} Our system is highly efficient and scalable: 
\emph{Result 1. Maximum load.} With high probability the maximum queue length is $O(\log \log n)$ at the stationary distribution, generalizing existing PoT results. \emph{Result 2. Learning speed.} The time to learn \emph{all the parameters $\mu_i$} is $O\left(\frac{\log (1/n)}{(1-\alpha)^2}\right)$. Thus, when the number of workers doubles, it takes only a constant amount of additional time to learn the system. 
\emph{Result 3. Convergence time.} When the system has reliable estimates of the processing powers, it takes additional $O(1)$ time to converge to the stationary distribution. 

\myparab{Interpreting the results.} 
Consider the ``life cycle'' of \sysname. 
At $t_1$, \sysname is in stationary distribution. We apply result 1: 
the maximum load at $t_1$ is $O(\log \log n)$ with high probability. At $t_2 > t_1$, the system experiences a shock (the processing power of a large number of workers get changed) and the estimates are no longer correct.
\sysname re-learns the new worker speeds. Result 2 states that the learning time is $O(\log 1/n)$. Let $t_3 = t_2 + O(\log 1/n)$ be the time the system re-learns the worker speeds. At this time, there could be backlogs during the learning (between time $t_2$ and $t_3$) since the system has been using inaccurate estimates. Result 3 states that it takes $O(1)$ time to clear up backlogs. 

\myparab{Notations.} We use $u, v$ to represent vectors, where $u_i$ ($v_i$) represents the $i$-th entry of $u$ ($v$). $\buv$ and $\bvv$ are random vectors. We refer \sysname as a PPoT process (\textbf{p}roportional sampling + PoT). We denote a PPoT process as $\{\buv(t)\}_t$. Let the stationary distribution of a PPoT be $\buv_{\pi}$. Abusing the notation, let $\{\buv_{\pi}(t)\}$ be a PPoT process that starts with stationary distribution. 
Below, we explain our analysis of the three key results. We start with result 2 because the analysis is more complex, followed by results 1 and 3.
See the technical report for the analysis~\cite{rosella-tr}).

\section{Evaluation}

\begin{figure}[t!]

 \includegraphics[scale=0.32]{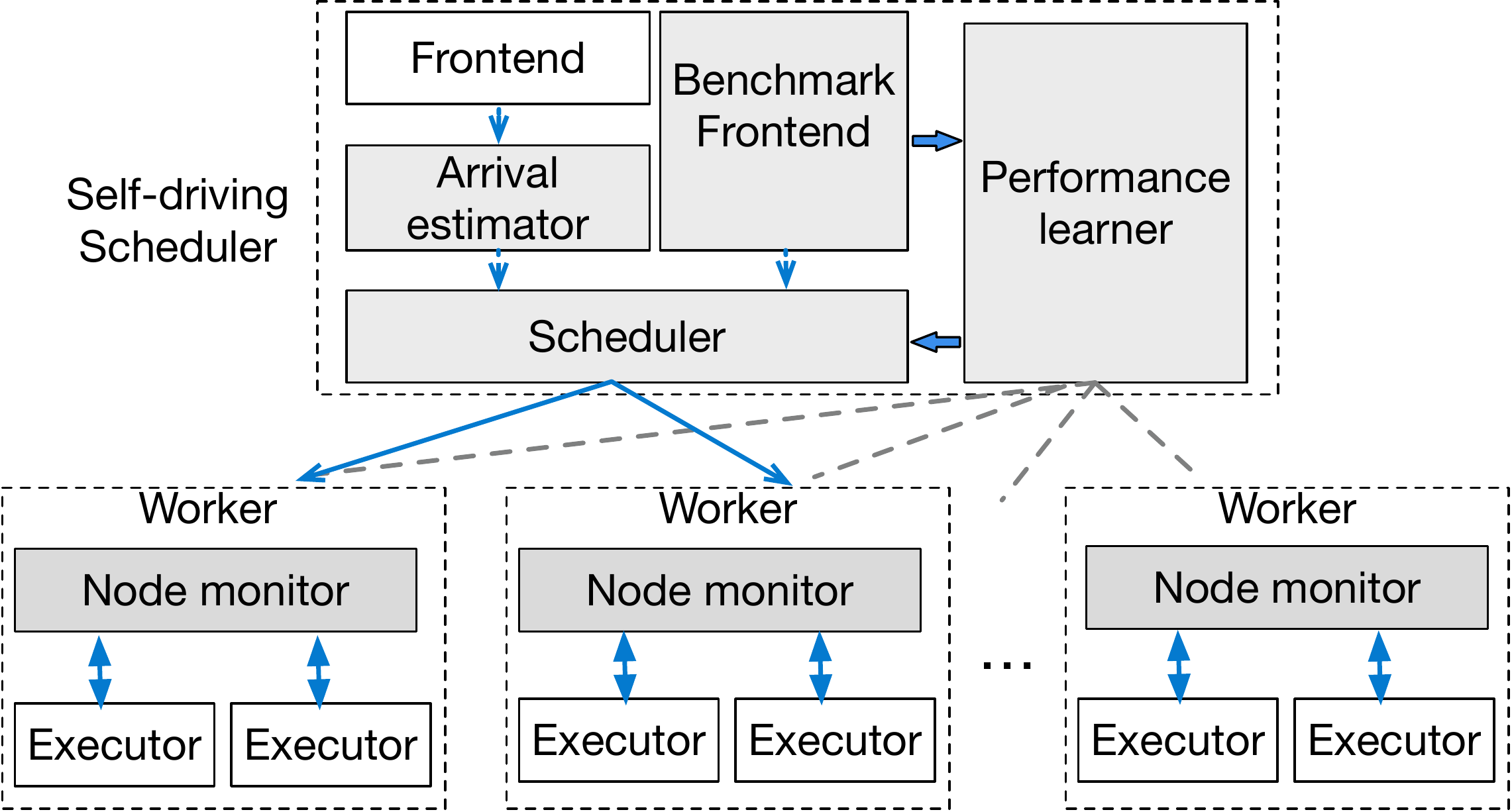}
  \centering

  \caption{Implementation of \sysname.}
    \label{fig:implementation}
\vspace{-4mm}
\end{figure}

\myparab{Implementation}
We implement (Fig.~\ref{fig:implementation}.) our key modules (arrival estimator, scheduling policy, and performance learner) on top of Spark with thrift handling the communication between components. 
Our modules are as follows. 
\mypara{At the frontend:} An arrival rate estimator estimates the load of the system.
A benchmark frontend dispatches jobs uniformly for the purpose of estimating worker speed.
A performance learner continuously estimates worker speed.

\mypara{At the backend:} Each node monitor maintains two separate queues: one queue is for the ``real'' jobs and the other is for the benchmark/fake jobs. When a benchmark or real task completes, the node monitor reports an updated estimation of worker speed to the performance learner.


\myparab{Setup} We perform our experiments in AWS clusters include 32 memory-optimized EC2 instances (m2.4xlarge, with 64G memory) consisting of one scheduler and 31 workers.
We examine \sysname for both real loads (TPC-H) and synthetic loads. Synthetic loads allow us to create extreme environments to understand the robustness of different systems. 

\myparab{Baselines} The baselines we examine include:  \emph{(i)} Power-of-two-choices, \emph{(ii)} Sparrow~\cite{ousterhout2013sparrow}, \emph{(iii)} PSS+Learning: it continuously estimates worker speeds and uses the estimates to run the proportional sampling algorithm, and \emph{(iv)} Multi-armed bandit: when a new job arrives, with $\eta$ probability, we uniformly choose a worker to serve the job. With $1- \eta$ probability, we use PSS+PoT.
\footnote{No existing system directly uses this baseline but since this is an intuitive design, we need to understand its performance.}. 
We examine $\eta \in\{ 0.2, 0.3\}$, and \emph{(v)} Halo~\cite{gandhi2015halo}: an heterogeneity-aware scheduler that assumes the knowledge of worker speeds. This will only be briefly examined in synthetic loads and we shall see its performance gain is limited even it has accurate knowledge of $\lambda$ and $\mu_i$'s. Other prior baselines (\eg uniform random, per task sampling, or batch sampling) are not included because their performances are significantly worse. 


\subsection{TPC-H Workload}\label{sec:exp_tpch}

The TPC-H benchmark~\cite{tpch} is representative of ad-hoc queries that have low latency requirements. Our experiments closely follow~\cite{ousterhout2013sparrow}. 
We use two query ids q3 and q6 in our experiments.

\mypara{Execution workflow.} The TPC-H workload is submitted to Shark, which compiles the queries into Spark stages (also referred to as requests) and waits for scheduling. 
\sysname resides inside Spark and controls the scheduling policy. 
Each stage corresponds to a job, which consists of multiple tasks. The benchmark consists of more than 32k tasks. Running the benchmark takes approximately 60 minutes. We report the result of the 40 minutes in the middle of execution (up to the point 30k tasks or 6.2k stages are completed). 

\mypara{Controlling worker speed.} While all EC2 instances are of the same type, we modify the executor in Spark to slow down a worker $k$ times: When a worker receives a task, it executes the task and records the execution time $T$. After completing the task, the worker holds the task $(k - 1)T$ more time and then informs to the node monitor that the task is completed.
The worker speeds ($\mu$'s) are from the set $\{0.01, 0.04, \dots 0.81\}$ to mimic heterogeneous environments. The choice of workloads are not critical. Other workloads also exhibit similar qualitative behaviors.

\mypara{Integration with late-binding (LB)~\cite{ousterhout2013sparrow}.} 
\sysname is compatible with late binding.

\myparab{Static environment.} We first consider worker speeds are known and do not change over time. The \emph{response time} of a job is the time between the job arrives at the scheduler and the time when the last task in the job is executed.

\begin{figure}

\centering
     \subfloat[Static environment.\label{subfig-1:dummy}]{%
       \includegraphics[width=0.35\textwidth]{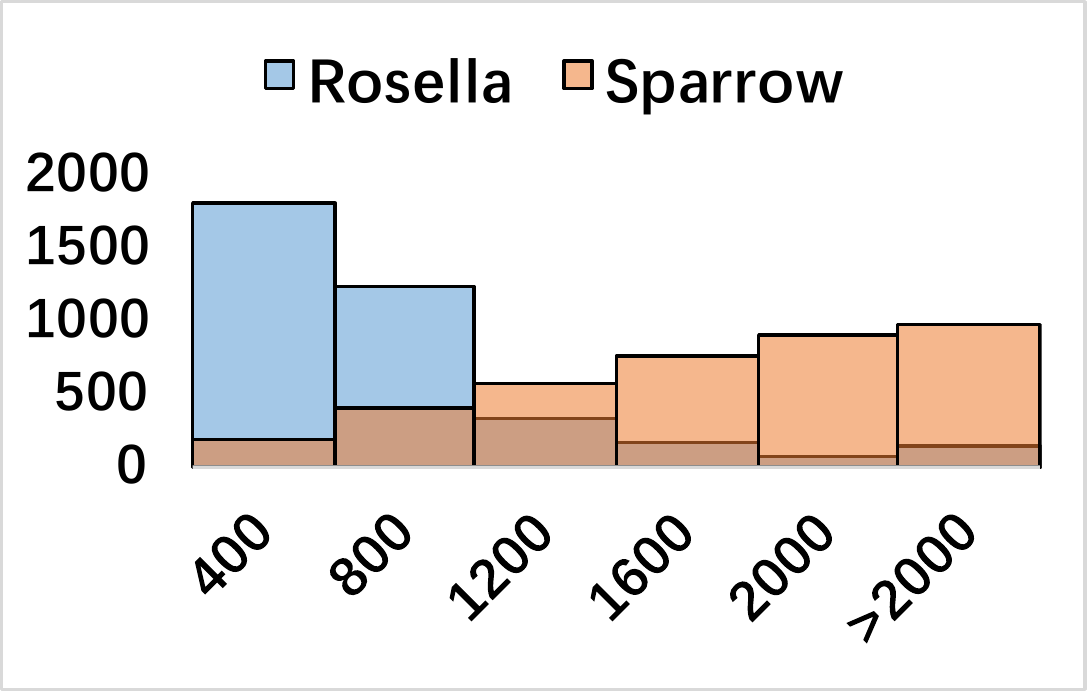}
     }
     \hspace{6mm}
     \subfloat[Volatile environments.\label{volatile_en}]{%
       \includegraphics[width=0.35\textwidth]{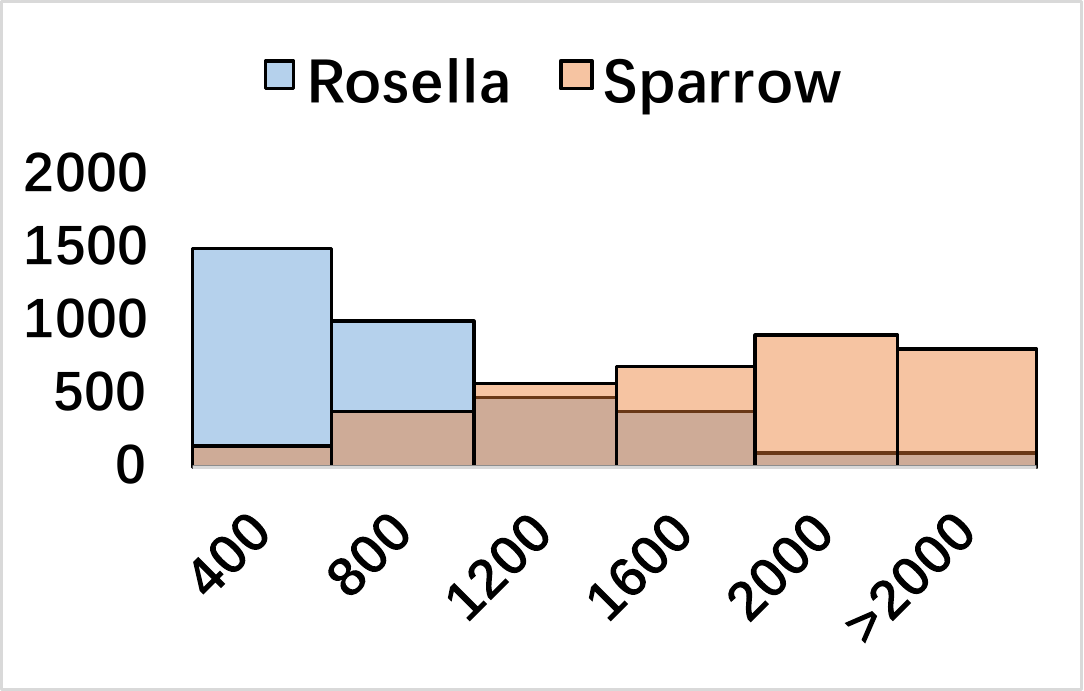}
       
     }
    \vspace{-2mm}
 \caption{{Distribution of response time for unconstrained requests.}}
     \label{fig:Histogram}
 \vspace{-5mm}
\end{figure}

Fig.~\ref{fig:Histogram}a shows the response time distributions of Sparrow and \sysname. \sysname's distribution decays exponentially before 2,000ms (thus, most jobs complete before 2,000ms) whereas Sparrow's distribution is \emph{monotonically increasing}, with a much larger portion of jobs that cannot be completed in 2,000ms. 

We next examine all the baselines. See the  Fig.7a. \sysname's performance is uniformly better than all other baselines in both q3 and q6.  
Multi-armed bandit algorithm has the worst performance. We can also see the breakdown of performance gains. Introduction of PSS helps the system to outperform Sparrow. When PoT and late-binding techniques are introduced, the performance continues to improve. The average response time for Sparrow is 1,901 while the average response time for \sysname is 675, corresponding to 65\% of improvement.

\begin{figure*}
\centering
   \includegraphics[width=0.56\textwidth]{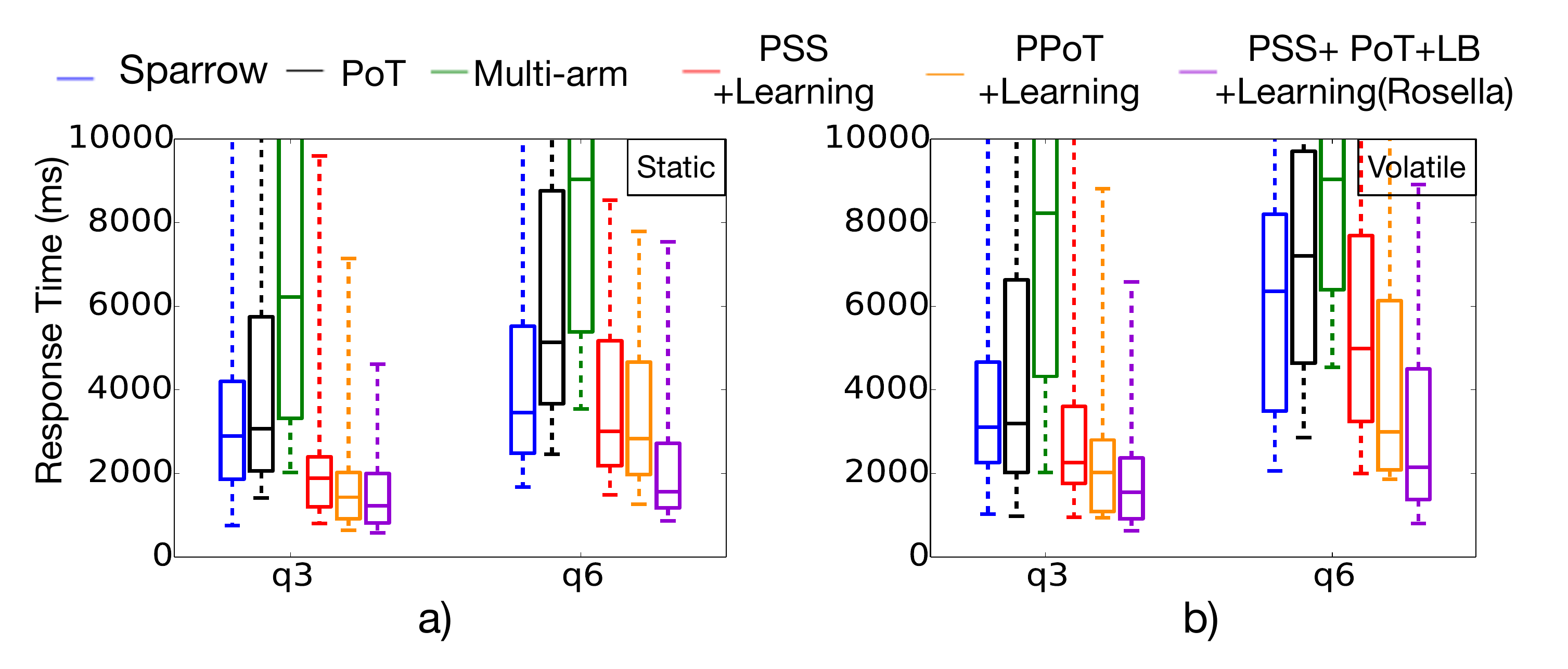}
  \vspace{-3mm}
 \caption{Response time (in ms) for TPC-H queries. The 5th, 25th, 50th, 75th and 95th percentiles are presented. The load is 0.8.}
     \label{fig:barchartstpch}
\end{figure*}

\myparab{Evolving worker speed.} We next consider the worker speeds evolve over time. We randomly permute the worker speeds every two minutes.
This setup ensures that the total throughput remains unchanged over time so that we can focus on the learning behaviors (instead of the overload behaviors) of the system.

Fig.~\ref{volatile_en} shows a comparison between \sysname and Sparrow. A significant portion of jobs at Sparrow cannot be completed with 2,000 seconds. The performance of \sysname degrades (compared to the static setting) because it needs to continuously learn the new worker speeds and adjust its scheduling policy. Sparrow's performance does not degrade  because it already oblivious to the worker speeds, and therefore changing the policy will not further ``harm'' sparrow. 

Fig.~\ref{fig:barchartstpch}b shows the performance of \sysname and all baselines.  \sysname has the best performance. PPS+Learning and PPoT+Learning all have better performance than Sparrow, PoT, or Multi-armed bandit. The algorithms that use learning (multi-armed, PSS+Learning, PPoT+Learning, \sysname) all have degraded performance compared to the static setting. The algorithms that do not learn (Sparrow and PoT) do not degrade for the same reason discussed above.

\section{Related Work}

\myparab{Schedulers.} The job scheduling problem is widely studied in industry and academia (\cite{hindman2011mesos,yarn,smith1978new,humphries2021case,kaffes2019centralized}). Mesos~\cite{hindman2011mesos}, YARN~\cite{yarn}, and Omega~\cite{schwarzkopf2013omega} are production-strength schedulers that allocate resources at coarse granularity. Because they need to support complex operations, they sacrifice request granularity and thus usually do not work for latency-sensitive tasks/jobs. A significant effort is devoted to design schedulers that are fair, often at the cost of reduced efficiency~\cite{isard2009quincy,wolf2010flex,ghodsi2011dominant,humphries2021case}.  

Performance-optimized schedulers~\cite{smith1978new,lin2013joint,ousterhout2013case} use a different set of techniques. For example, the ``shortest remaining processing time'' policy prioritizes smaller jobs so that the average waiting time is optimized~\cite{smith1978new,lin2013joint,moseley2011scheduling,wang2013preemptive}. These techniques are not applicable in our setting because our jobs are relatively homogeneous, and all jobs have low latency requirements. The so-called ``late-binding'' technique formally introduced in~\cite{ousterhout2013sparrow} also resembles an earlier technique developed by Dean~\cite{Dean2013}. Roughly speaking, the system may send the same requests to multiple workers but cancel the remaining outstanding requests while one of them is processed/completed. Ananta ~\cite{patel2013ananta} is a layer-4 load-balancer that combines techniques in networking and distributed systems to refactor its functionality to meet scale, performance, and reliability requirements.  

Distributed schedulers
focus on a system's scalability and are often designed to minimize coordination/communication. Duet~\cite{gandhi2015duet} is a distributed hybrid load balancer that fuses switch with the software load balancers. Apollo~\cite{boutin2014apollo} is a coordinated scheduling framework that is suitable for executing jobs requiring heavy resources. 

\myparab{Theory.} Load balancing algorithms and PoT in homogeneous systems have been extensively studied. See~\cite{Mitzenmacher2005} for a comprehensive treatment of balls-and-bins, ~\cite{Mitzenmacher00thepower} for a survey of PoT algorithms in discrete-time systems, and~\cite{ying2017power,xie2015power} for more recent developments. For the continuous-time counterpart, see \cite{McDiarmid05,Bramson2010} and references therein. Halo~\cite{gandhi2015halo} provides an optimal scheduling policy in heterogeneous environments when the speed of the workers are known, and the scheduler can only probe one machine. 
\cite{mukhopadhyay2016analysis} studies a similar model to ours but they assume a constant number of worker types (the number of distinct $\mu_i$'s is $O(1)$), and their speeds are known. Online estimation and change point detections are an extensively studied area~\cite{poor2013introduction, tartakovsky2014sequential}. For recent development in multi-armed bandit algorithms, see~\cite{bubeck2012regret}.
Stochastic optimization and exponential moving average~\cite{harold1997stochastic} are widely used techniques to estimate the means of a sequence of i.i.d. random variables. 
They perform one proportional sampling at the group level (workers with the same speed are in the same group), and perform a PoT inside the group. This algorithm cannot be directly generalized to our setting. 
Online estimation and change point detections are an extensively studied area~\cite{poor2013introduction, tartakovsky2014sequential}. 
Stochastic optimization and exponential moving average~\cite{harold1997stochastic} are widely used techniques to estimate the means of a sequence of i.i.d. random variables. For recent development in multi-armed bandit algorithms, see~\cite{bubeck2012regret}.

\section{Conclusion}

This paper introduces \sysname, a scalable self-driving scheduler for heterogeneous and volatile environments. \sysname achieves high throughput and low latencies by introducing two key modules: the scheduling policy leverages proportional sampling and power-of-two-choices to optimize the queue length, and the performance learner introduces benchmark/fake jobs and uses a dynamic sliding window to achieve optimal learning strategy.  
Our experiments show that \sysname significantly outperforms prior algorithms, and is robust against various workloads. 

{\scriptsize
\bibliographystyle{unsrt}
\bibliography{relatedwork}
}

\ifpaper
\else
\appendix 
\section{Missing proofs}
\begin{proof}[Proof of Lemma~\ref{lem:ppotgood2}]
Recall that $\epsilon = \frac 3 {10}(1-\alpha)$. Let $\buv(t)$ be the process using the estimates $\hat \mu_i$ in the $\sqq$ policy. Let us construct another process $\buv'(t)$ so that the true processing power of worker $i$ is $\hat \mu_i$. This process also uses $\hat \mu_i$ as its estimate of worker $i$ (\ie its estimates are accurate). 

Observe that we may couple two processes such that \emph{(i)} both processes will choose the same set of candidate workers upon the arrival of a new job, and \emph{(ii)} Because we guarantee that $\hat \mu_i$'s are underestimates of $\mu_i$, $\buv'$ is ``strictly slower'' than $\buv$ (\ie for any fixed $i$, worker $i$ is slower in $\buv'$ than that in $\buv$). 

By induction, we can prove that $\buv(t) \preceq \buv'(t)$. Also, we can also see that $\sum_{i \leq n}\hat \mu_i > \lambda$, which implies $\buv'$ is a stationary process. Therefore, we know that $\buv$ is stationary, has max queue length $O(\log \log n)$, and has $O(1)$ convergence time. 
\end{proof}
\fi


\end{document}